\documentclass[apsamssymb,amsmath,,twocolumn,secnumarabic]{revtex4}

\newcommand{\QZ}{Z}
\newcommand{\xib}{\xi^b} 
\newcommand{\xic}{\xi^c}

\newcommand\id{{\text{id}}}
\newcommand\tr{{\text{tr}}}

\begin{document}

\title{New boundary conditions for the $c=-2$ ghost system}

\author{Thomas Creutzig, Thomas Quella,
 Volker Schomerus}

\affiliation{
DESY Theory Group, DESY Hamburg, Notkestrasse 85, D-22603 Hamburg,
Germany}

\affiliation{KdV Inst. for Mathematics, Univ. of Amsterdam,
Plantage Muidergracht 24, 1018 TV Amsterdam, The Netherlands}

\affiliation{Center for Mathematical Physics, Bundesstrasse 55,
D-20146 Hamburg, Germany}

\affiliation{King's College London, Department of Mathematics, 
Strand, London WC2R 2LS, United Kingdom} 

\date{December 2006}

\begin{abstract}
We investigate a novel boundary condition for the $bc$ system with
central charge $c=-2$. Its boundary state is constructed and tested 
in detail. It appears to give rise to the first example of a local 
logarithmic boundary sector within a bulk theory whose Virasoro zero 
modes are diagonalizable. 
\vskip 0.1cm
\hskip -0.3cm
PACS numbers: 11.25. HF; 11.25.-w; 11.25. Sq.
\end{abstract}

\maketitle

\vspace*{-7.5cm} \noindent
{\tt DESY 06-219  \hspace*{12.7cm} NSF-KITP-06-122}\\
{\tt KCL-MTH-06-15 \hspace*{12.5cm} hep-th/0612040}\\[5cm] 
 
\section{Introduction} 
The $bc$ system with Virasoro central charge $c=-2$, and the 
closely related symplectic fermion model, have been studied 
extensively in the past, both in the bulk and on the boundary
(see e.g.\ \cite{Kausch:2000fu,Gaberdiel:2001tr,Flohr:2001zs,%
Gaberdiel:2006pp} and references therein). 
Most of the past work was driven by formal questions in the 
context of logarithmic conformal field theory. Recently, it 
was pointed out \cite{Schomerus:2005bf,Gotz:2006qp,Quella:2007} 
that the $bc$ system at $c=-2$ also plays a crucial role for the 
solution of WZNW models on supergroups. In fact, it enters through 
a Kac-Wakimoto type representation of such theories. The latter 
reduces the solution of the WZNW model on superspaces to that 
on the corresponding bosonic base. In order to extend such a 
free fermion representation to the boundary sector, we have 
to impose boundary conditions on the $bc$ system.  
Surprisingly, it turns out that the relevant Neumann-type 
boundary theory has not been discussed in the existing 
literature. We shall fill this gap below. 

As the name indicates, the $bc$ system involves two sets of 
chiral bulk fields $c,\bar c$ and $b,\bar b$ of conformal
dimension $h_c = 0$ and $h_b = 1$, respectively. In the
conventional setup, we would glue $c$ to $\bar c$ and $b$
to $\bar b$ along the boundary \cite{Callan:1987px}. But for $c=-2$ 
there exists another possibility: namely, to glue $b$ to a derivative 
of $\bar c$ and  vice versa. More precisely, we can demand that
\begin{equation} \label{bc}
b(z) \, = \, \mu \bar \partial \bar c(\bar z) \  \ , \  \
\bar b(\bar z) \, = \, - \mu \partial c(z) \  \  \text{ for }
  \  z  = \bar z\ \ .
\end{equation}
These relations guarantee trivial gluing conditions for the
energy momentum tensor $T=-b\partial c$.
It is not difficult to check that the action of the $bc$ system
is invariant under variations respecting (\ref{bc}) provided
we add an appropriate boundary term,
\begin{equation} \label{act}
S \ = \ \frac{1}{2\pi} \int d^2z \left[ b\,\bar \partial c +
           \bar b\,\partial\bar c \right]
  - \frac{i\mu}{4\pi} \int du \ c\,\partial_u \bar c \ \ .
  \end{equation}
Our aim here is to solve the theory that is defined by
the action (\ref{act}) and the boundary condition
(\ref{bc}). We shall set $\mu =1$ throughout our
discussion. Formulas for the general case are easily
obtained from the ones we display below.

\section{Solution of the boundary theory} 

In order to construct the state space and the fields
explicitly, we introduce an algebra that is generated
by the modes $c_n, b_n$ and two additional zero
modes $\xib_0, \xic_0$ subject to the conditions
\begin{eqnarray} \label{comm}
\{ c_n,b_m\} & = & n\,\delta_{n,-m} \ \ ,\\[2mm]
 \{ \xic_0,b_0 \} \ = \ 1 \ \ \ \ & ,  & \ \ \ \ %
 \{\xib_0 , c_0 \} \ = \ 1 \ \ .
\end{eqnarray}
All other anti-commutators in the theory are assumed to
vanish. The state space of our boundary theory is
generated from a ground state with the properties
\begin{equation}\label{vac}
c_n | 0\rangle \ = \ b_n |0\rangle\ = \ 0
\ \ \ \text{ for } \ \ \ n \ \geq \ 0 
\end{equation}
by application of `raising operators', including
the zero modes $\xib_0$ and $\xic_0$. On this
space we can introduce the local fields $c,\bar c,
b,\bar b$ through the prescription
\begin{eqnarray} \label{b6} 
b(z) &= & \ \sum_{n\in\QZ} b_n z^{-n-1} \\[2mm]
c(z) & = & \sum_{n\neq0} \frac{c_n}{n} z^{-n} +
 c_0 \ln z + \xic_0\\[2mm]
\bar b(\bar z) & = & \sum_{n\neq 0} c_n \bar z^{-n-1} - c_0 \bar
z^{-1}
\\[2mm]
\bar c(\bar z)  & = & - \sum_{n\neq 0} \frac{b_n}{n} \bar z^{-n} +
b_0 \ln \bar z - \xib_0 \label{bc9} 
\end{eqnarray}
It is not difficult to check with the help of
eqs.\ (\ref{comm}) that these fields satisfy
the correct local anti-commutation relations
$$
\bigl\{ b(z), c(w) \bigr\} \ = \ \delta(z-w) \ \ , \ \
\bigl\{ \bar b(\bar z),\bar c(\bar w) \bigr\} \ = \
\delta(\bar z- \bar w)\ \
$$
in the interior of the upper half plane.
Needless to stress that they also fulfill our
boundary conditions (\ref{bc}) with $\mu = 1$.

For later use let us also spell out the construction
of the Virasoro generators in terms of fermionic modes,
$$ L_n \ = \  \sum_{m \neq 0} \,  :b_{n-m} c_m : - b_n
    c_0\ \ . $$
It is important to stress that -- due to the term $c_0 b_0$
--  the element $L_0$ satisfies $L_0 \xic_0\xib_0 |0\rangle = 
|0\rangle$. Since $L_0$ vanishes on all other ground states, it 
is non-diagonalizable. In other words, our boundary theory is an
example of a logarithmic conformal field theory. The logarithms in
this model, however, are restricted to the boundary sector since
the Hamiltonian of the bulk theory is diagonalizable (see below).

Computations of correlation functions in our boundary theory, 
require to introduce a dual vacuum with the properties
\begin{eqnarray}\label{cav}
\langle 0| c_n \ = \ \langle 0| b_n  & = & 0
\ \ \  \text{ for }  \ \ \ n \ \leq \ 0 \\[2mm]
\langle 0 | \xic_0 \xib_0 |0\rangle 
& = &  1 \ \ \ \ .  \label{cavn} 
\end{eqnarray}
 For the c=-2 ghost system the ground states $|0\rangle$ and $\langle 
0|$ which are annihilated by the zero modes $b_0$ and $c_0$ are at the 
same time $SL(2,C)$ invariant vacua. Consistency with the commutation 
relations requires $\langle 0|0\rangle = \langle 0|\{b_0,\xic_0\}
|0\rangle =0$. Therefore, the simplest non-vanishing quantity is 
$\langle 0|\xic_0\xib_0|0\rangle$ for our boundary theory (see also 
\cite{Flohr:2001zs} for a more detailed discussion).

Finally, we would like to display the boundary state $|N\rangle$ 
for our new boundary condition. Before we provide explicit 
formulas let us briefly recall that the bulk fields are 
obtained as 
$$ 
c(z) \ =  \ \xic_0 + \sum_{n \neq 0} \, \frac{c_n}{n} \, z^{-n}  
\ \ \ , \ \ \ b(z) \ =  \ \sum_{n\in Z} \, b_n \, z^{-n-1} 
$$
and similarly for their anti-holomorphic counterparts. Note that 
there are no modes $c_0,\bar c_0$ and $\xib_0,\bar\xib_0$ in the
bulk of our $bc$ ghost system. This feature distinguishes the 
$c=-2$ ghosts from the closely related symplectic fermions. 
According to the standard rules, the boundary state for our 
boundary theory must satisfy the following Ishibashi conditions 
\cite{Ishi}
\begin{equation} \label{Ish1} 
(b_n -  \bar c_{-n}) |N\rangle \ = \ 0 \ \ , \ \ (c_n + \bar
b_{-n}) |N\rangle \ = \ 0 \ \
\end{equation}
for $n\neq 0$ and $b_0 |N\rangle = \bar b_0 |N\rangle = 0$. As
one may easily check, the unique solution to these conditions 
is given by
\begin{equation}\label{BS}
\frac{|N\rangle}{\sqrt{2\pi}} \, = \, 
\exp\left(-\sum_{m=1}^\infty\,
 \frac{1}{m} ( c_{-m} \bar c_{-m} +
  b_{-m} \bar b_{-m}) \right) |0\rangle
\end{equation}
where $|0\rangle$ is a state in the bulk theory that satisfies
conditions of the form (\ref{vac}) for both chiral and anti-chiral
modes. There also exists a dual boundary state $\langle N|$,
satisfying the conditions
\begin{equation}
\langle N| (b_n + \bar c_{-n}) \ = \ 0 \ \ , \ \ \langle N | (c_n
- \bar b_{-n}) \ = \ 0 \ \
\end{equation}
for $n\neq 0$ and $\langle N| b_0  = \langle N| \bar b_0 = 0$.
These linear relations are related to eqs. (\ref{Ish1}) by 
conjugation using that $c_n^* = - c_{-n}$ for $n \neq 0$ and 
$b^*_n = b_{-n}$ etc. The dual boundary state is given by the 
following explicit formula
\begin{equation}\label{dBS}
\frac{\langle N|}{\sqrt{2\pi}} \, = \, 
\langle 0|\, \exp\left(\sum_{m=1}^\infty \, 
  \frac{1}{m}\, ( c_{m} \bar c_{m} +
    b_{m} \bar b_{m}) \right)
\end{equation}
involving a dual closed string ground state $\langle 0|$ that
obeys conditions of the form (\ref{cav}) for modes of chiral and
anti-chiral fields and that is normalized by $\langle 0 | 
\xic_0 \bar \xic_0 |0\rangle=1$ (see comments after eq.\ 
\ref{cavn}).   

\section{Cardy consistency conditions} 

Having constructed our new boundary theory, and in 
particular its boundary state, we would now like to 
perform two Cardy-like consistency tests. To begin 
with, let us verify that $|N\rangle$ 
satisfies world-sheet duality. We stress that in 
this note we consider a theory in which bulk and 
boundary theory consist of Ramond sectors only, a 
choice that we shall comment in more detail below. 
In such a model, world-sheet duality relates quantities
that are periodic in both world-sheet space and
time. The simplest such quantity in our boundary
theory would be $\tr[q^{L_0 + 1/12} (-1)^F]$ which
vanishes since bosonic and fermionic states come
in pairs on each level of the state space. The
same is certainly true for $\langle N| \tilde q^{L_0  + 
1/12} (-1)^{F} |N\rangle$, in agreement 
with world-sheet duality. In order to probe finer 
details of the theory, we need to 
consider quantities with additional insertions 
of fields or zero modes. Here, we shall establish 
the relation
\begin{eqnarray}\nonumber
& & \tr \left(q^{H^o} (-1)^F c(z) \bar c(\bar z)  \right)  \ = \ 
 \\[2mm] & & \hspace*{1cm} \ = \ \langle N|
 \tilde q^{\frac12 H^c} (-1)^{\frac12 F^c} c(\xi) \bar 
  c(\bar \xi)  |N\rangle \label{test1}
\end{eqnarray}
where $H^o = L_0 + 1/12, q = \exp(2\pi i \tau), \xi = 
\exp(-\frac{1}{\tau}\ln z)$ and $F^c = F + \bar F$, as usual. 
The closed string Hamiltonian is given by
$$
H^c \ = \  \sum_{m\in\QZ}\Bigl[:b_{-m} c_m: + :\bar b_{-m} 
\bar c_m:\Bigr]   + 1/6 \ \ .
$$
Validity of eq.\ (\ref{test1}) is required by the definition 
of boundary states (see e.g.\ \cite{Recknagel:1997sb}).   
Starting with the left hand side, it is rather easy to 
see that
\begin{eqnarray} 
 \tr \left(q^{L_0 + 1/12} (-1)^F c(z) \bar c(\bar z)
 \right)\!\!&=&\!\!-\tr\left(q^{L_0 + 1/12} (-1)^F \xic_0 \xib_0
 \right) \nonumber \\[2mm] & & \hspace*{-3cm} \ = \  
  2 \pi i \tau \eta(q)^2 \ = \ - 2 \pi \eta(\tilde q)^2\ \ . 
\end{eqnarray} 
In the computation we split off the
term $c_0 b_0$ from $H^o$ and use it to saturate the
fermionic zero modes. The rest is then straightforward. We can
reproduce the same result if we insert our explicit formulas for
the boundary states $|N\rangle$ and $\langle N|$ into the right
hand side of eq.\ (\ref{test1}).

It is possible to perform another similar test of our
boundary theory using the usual trivial boundary conditions
of the ghost system. In this case, the field $c(z)$ is
identified with its own anti-holomorphic partner $\bar
c (\bar z)$ along the boundary and likewise for
the pair $b$ and $\bar b$. Let us recall that the boundary
state $|\id\rangle $ and its dual $\langle \id |$ take the
form \cite{Callan:1987px}
\begin{eqnarray} \nonumber
    |\id \rangle & = &
 \text{exp}\Biggl(\sum_{m=1}^\infty \frac{1}{m} \Bigl(c_{-m}\bar{b}_{-m} +
  \bar{c}_{-m} b_{-m}\Bigr)\Biggr)(\xic_0-\xi^{\bar c}_0)|0\rangle
  \\[2mm] \langle \id |&=&i\langle0|(\xic_0-\xi^{\bar c}_0)\,\text{exp}
 \Biggl(\sum_{m=1}^\infty\, \frac{1}{m} \Bigl(\bar{b}_{m}c_{m}+b_{m}
 \bar{c}_{m}\Bigr)\Biggr)
\end{eqnarray}
where we use the same notations as before. For the exchange 
of closed string modes between $|N\rangle$ and $\langle \id |$ 
the above formulas imply
\begin{eqnarray}
        \langle\id |\,  \tilde q^{\frac12 H^c }\,  (-1)^{\frac12 F^c}\,
c(\xi) \, |N\rangle & = &  \langle\id |\,  
  \tilde q^{\frac12 H^c }\,  (-1)^{\frac12 F^c}\, \xic_0 \, |N\rangle
\nonumber \\[2mm]
 & & \hspace*{-4cm} \ = \  
\sqrt{2\pi}\,  \tilde{q}^{\frac{1}{12}}\prod_{n=1}^\infty
\bigl(1+\tilde{q}^{2n}\bigr)
\ = \
\sqrt{\frac{\pi \theta_2(2\tilde{\tau})}{\eta(2\tilde{\tau})}}\ \ .
\label{res1}
\end{eqnarray}
Once more we had to insert the field $c(z)$ in order to get
a non-vanishing result. For comparison with a world-sheet dual,
we need to quantize the ghost system 
on the upper half plane with trivial boundary conditions on the
positive real axis and our non-trivial ones on the other half. 
A moment of reflection reveals that the following
combinations 
\begin{eqnarray*} 
\chi^+(z) & = & \frac{1}{\sqrt{2}}\bigl(b(z)+i\partial c(z)\bigr) \ \ , 
\\[2mm]  
\chi^-(z) & = & \frac{1}{\sqrt{2}} \bigl(ib(z)+\partial c(z)\bigr)
\end{eqnarray*} 
diagonalize the monodromy, i.e. they obey the following simple 
periodicity relations $\chi^\pm(e^{2\pi i} z) = \pm i 
\chi^\pm(z)$. Hence, they take the following form  
\cite{Kausch:2000fu}
\begin{eqnarray*}
\chi^\pm(z) &= &\sum_{r\in\QZ\mp\frac{1}{4}}\chi_r^\pm\,z^{-r-1}\ \ .
\end{eqnarray*}
The modes $\chi^\pm_r$ satisfy the same canonical commutation
relations, $\{\chi^+_r,\chi^-_s\}=r \delta_{r,-s}$, as before.
Formulas for the Virasoro generators can easily be worked out. 
For us, it suffices to display the zero mode $\tilde L_0$,
\begin{equation}
 \tilde L_0 \ = \ -\sum_{r\in\QZ-\frac{1}{4}}:\chi^+_r\chi^-_{-r}:- 
\frac{3}{32}\ \ .
\end{equation}
The constant shift by $3/32$ is needed in order to obtain
standard Virasoro relations with the other generators (see 
also \cite{Saleur:1991hk} for a closely related analysis 
of twisted sectors in the bulk theory). The state space of 
our boundary theory contains two ground states $|\Omega_\pm
\rangle$ which are related to each other by the action of a
zero mode $\xic$. On this space we can introduce the 
field $c$ through 
$$ c(z) \ = \ \sqrt{\pi} \xic + \frac{i}{\sqrt2} 
\sum _{r\in\QZ -\frac{1}{4}}\frac{\chi_r^+}{r} \,z^{-r}
- \frac{1}{\sqrt2} \sum _{r\in\QZ+\frac{1}{4}}\frac{\chi_r^-}{r} 
\,z^{-r}\ . 
$$
From the construction of the state space and our formula for
$\tilde H^o = \tilde L_0 + 1/12$ we infer the following
expression for the mixed open string amplitude,
\begin{eqnarray}
    \tr\Bigl(q^{\tilde H^o} (-1)^F c(z) \Bigr) & = &
 \nonumber \\[2mm]
 & & \hspace*{-3cm} \ = \ \sqrt{\pi}\,  q^{-\frac{1}{96}}
\prod_{n=0}^\infty\Bigl(1-q^{\frac{1}{2}(n+1/2)}\Bigr)\ =\
\sqrt{\frac{\pi \theta_4(\tau/2)}{\eta(\tau/2)}},
\nonumber
\end{eqnarray}
which reproduces exactly the previous result (\ref{res1})
upon modular transformation and concludes our investigation 
of the new boundary theory. Let us remark that the same 
partition function was found recently in \cite{Jacobsen:2006bn} 
with the help of boundary loop models. 

\section{Conclusions and outlook}  

The choice of our new gluing condition for the $bc$ system 
was motivated by the interest in branes on supergroups. As we 
shall discuss in a forthcoming paper, maximally symmetric branes 
in a WZW model on a supergroup turn out to satisfy Neumann-type 
boundary conditions in the fermionic coordinates. This implies 
that all fermionic zero modes must act non-trivially on the 
space of open string states. In our toy model, the role of the 
fermionic coordinates is played by $c$ and $\bar c$. Hence, we 
needed to find boundary conditions with a four-fold degeneracy
of ground states. For the standard boundary conditions of the 
$bc$ system, $c = \bar c$ along the boundary and hence only one 
fermionic zero mode survives, giving rise to a 2-dimensional 
space of ground states. In this sense, the usual boundary 
conditions of the $bc$ systems are localized in one of the 
fermionic directions. Our boundary conditions come with two 
non-vanishing zero modes $\xib_0$ and $\xic_0$ (and their dual
momenta $c_0$ and $b_0$). This property makes them a  good 
model for maximally symmetric branes on supergroups.

There exist various extensions of our theory that we want to 
briefly comment about. In our analysis we focused on the RR 
sector of the $bc$ ghost system in the bulk. It 
is certainly straightforward to include a NSNS sector in case 
this is required by the application. Furthermore, we can also 
replace the bulk theory by its logarithmic cousin, the symplectic 
fermion model. Since the formulas and results are very 
similar, we refrain from giving more details. Boundary theories 
for symplectic fermion theories have  been studied extensively 
in the past (see e.g.\ \cite{Moghimi-Araghi:2000cx,Kogan:2000fa,%
Ishimoto:2001jv,Kawai:2001ur,Bredthauer:2002ct,Kawai:2002fu,%
Bredthauer:2002xb,Gaberdiel:2006pp}). We would like to stress, 
however, that our boundary condition seems to be new, also 
in the context of symplectic fermions. 

Let us be a bit more specific and relate our constructions 
to the results in \cite{Gaberdiel:2006pp}. A comparison of 
the gluing conditions shows that our state $|\id\rangle$ is a 
close relative of the $(N,\pm)$ boundary condition of 
\cite{Gaberdiel:2006pp}. In fact, all boundary theories 
considered in \cite{Gaberdiel:2006pp} 
glue $\partial c $ to $\bar \partial\bar c$ and $b$ to 
$\bar b$, with different choices of signs. None of these  
models displays any enhancement of zero-modes in the 
boundary spectrum. In this sense, we would prefer to 
consider them all as being of the same type (mixed 
Dirichlet-Neumann in the context of the $bc$ system). 
Using the notations of \cite{Gaberdiel:2006pp}, our new 
boundary theories arise when we glue $\chi^+$ to $\bar 
\chi^-$ and vice versa. Such a choice gives rise to a 
non-trivial gluing automorphism on the so-called triplet 
algebra and therefore it was excluded from the analysis 
in \cite{Gaberdiel:2006pp}. 

In the case of the $bc$ ghost system, the boundary state 
$|N\rangle$ has a rather novel feature: it describes a logarithmic 
boundary theory in a non-logarithmic bulk. Put differently, the 
$bc$ ghost system possesses a diagonalizable bulk Hamiltonian 
$H^c$. Nevertheless, the Hamiltonian $H^o$ of our new boundary 
theory is non-diagonalizable. Hence, logarithmic singularities 
can appear, but {\em only} when two boundary fields approach 
each other. To the best of our knowledge, such a behavior has 
never been encountered before. It shows that conformal field 
theories may be logarithmic even if none of its correlators 
on the sphere contain logarithms. 

\begin{acknowledgments}
We thank Matthias Gaberdiel, Andreas Recknagel, Sylvain Ribault,
Ingo Runkel and Hubert Saleur for their useful comments and
critical remarks.  TQ acknowledges the warm hospitality at 
the KITP in Santa Barbara during the completion of this article.
This work was partially supported by the EU Research Training 
Network grants ``Euclid'', contract number HPRN-CT-2002-00325, 
``Superstring Theory", contract number MRTN-CT-2004-512194, and 
``ForcesUniverse'', contract number MRTN-CT-2004-005104, as well 
as by the PPARC rolling grant PP/C507145/1 and by the National 
Science Foundation under Grant No. PHY99-07949. Until September 
2006 TQ has been funded by a PPARC postdoctoral
fellowship under reference PPA/P/S/2002/00370.
\end{acknowledgments}





\widetext


\begin{thebibliography}{99}

\bibitem{Kausch:2000fu}
  H.~G.~Kausch,
  {\it Symplectic fermions},
  Nucl.\ Phys.\ B {\bf 583} (2000) 513.

\bibitem{Gaberdiel:2001tr}
  M.~R.~Gaberdiel,
  {\it An algebraic approach to logarithmic conformal field theory},
  Int.\ J.\ Mod.\ Phys.\ A {\bf 18} (2003) 4593. 

\bibitem{Flohr:2001zs}
  M.~Flohr,
  {\it Bits and pieces in logarithmic conformal field theory},
  Int.\ J.\ Mod.\ Phys.\ A {\bf 18} (2003) 4497. 

\bibitem{Gaberdiel:2006pp}
  M.~R.~Gaberdiel and I.~Runkel,
  {\it The logarithmic triplet theory with boundary},
  hep-th/0608184.



\bibitem{Schomerus:2005bf}
  V.~Schomerus and H.~Saleur,
  {\it The GL(1$|$1) WZW model: From supergeometry to logarithmic CFT},
  Nucl.\ Phys.\ B {\bf 734} (2006) 221.

\bibitem{Gotz:2006qp}
  G.~G\"otz, T.~Quella and V.~Schomerus,
  {\it The WZNW model on PSU(1,1$|$2)},
  arXiv:hep-th/0610070.

\bibitem{Quella:2007}
  T.~Quella and V.~Schomerus, in preparation 

\bibitem{Callan:1987px}
  C.~G.~.~Callan, C.~Lovelace, C.~R.~Nappi and S.~A.~Yost,
  {\it Adding holes and crosscaps to the superstring}, 
  Nucl.\ Phys.\ B {\bf 293} (1987) 83.

\bibitem{Ishi} N.Ishibashi, {\it The boundary and crosscap states
in conformal field theories}, Mod. Phys. Lett. {\bf A 4} (1989)
251.





\bibitem{Recknagel:1997sb}
  A.~Recknagel and V.~Schomerus,
  {\it D-branes in Gepner models},
  Nucl.\ Phys.\ B {\bf 531} (1998) 185
  [arXiv:hep-th/9712186].

\bibitem{Saleur:1991hk}
  H.~Saleur,
  {\it Polymers and percolation in two-dimensions and 
  twisted N=2 supersymmetry},
  Nucl.\ Phys.\ B {\bf 382} (1992) 486
  [arXiv:hep-th/9111007].

\bibitem{Jacobsen:2006bn}
  J.~L.~Jacobsen and H.~Saleur,
  {\it Conformal boundary loop models},
  arXiv:math-ph/0611078.

\bibitem{Moghimi-Araghi:2000cx}
  S.~Moghimi-Araghi and S.~Rouhani,
  {\it Logarithmic conformal field theories near a boundary},
  Lett.\ Math.\ Phys.\  {\bf 53} (2000) 49
  [arXiv:hep-th/0002142].

\bibitem{Kogan:2000fa}
  I.~I.~Kogan and J.~F.~Wheater,
  {\it Boundary logarithmic conformal field theory},
  Phys.\ Lett.\ B {\bf 486}, 353 (2000)
  [arXiv:hep-th/0003184].

\bibitem{Ishimoto:2001jv}
  Y.~Ishimoto,
  {\it Boundary states \ in \ boundary logarithmic CFT},
  Nucl.\ Phys.\ B {\bf 619}, 415 (2001)
  [arXiv:hep-th/0103064].

\bibitem{Kawai:2001ur}
  S.~Kawai and J.~F.~Wheater,
  {\it Modular transformation and boundary states in logarithmic conformal  field
  theory},
  Phys.\ Lett.\ B {\bf 508}, 203 (2001)
  [arXiv:hep-th/0103197].

\bibitem{Bredthauer:2002ct}
  A.~Bredthauer and M.~Flohr,
  {\it Boundary states in c = -2 logarithmic conformal field theory},
  Nucl.\ Phys.\ B {\bf 639}, 450 (2002)
  [arXiv:hep-th/0204154].

\bibitem{Kawai:2002fu}
  S.~Kawai,
  {\it Logarithmic conformal field theory with boundary},
  Int.\ J.\ Mod.\ Phys.\ A {\bf 18}, 4655 (2003)
  [arXiv:hep-th/0204169].

\bibitem{Bredthauer:2002xb}
  A.~Bredthauer,
  {\it Boundary states and symplectic fermions},
  Phys.\ Lett.\ B {\bf 551} (2003) 378
  [arXiv:hep-th/0207181].

\end{thebibliography}
\end{document}